\documentclass[twocolumn,showpacs,preprintnumbers,amsmath,amssymb]{revtex4}

\usepackage{graphicx}
\usepackage{dcolumn}
\usepackage{bm}
\usepackage{subfigure}
\usepackage{epstopdf}
\usepackage{wrapfig}

\begin{document}

\preprint{APS/123-QED}

\title{The Role of Adhesion in the Mechanics of Crumpled Polymer Films}

\author{Andrew B. Croll}
\email{andrew.croll@ndsu.edu}
\affiliation{Department of Physics and Materials and Nanotechnology Program, North Dakota State University}%
\author{Timothy Twohig}
\affiliation{Department of Physics, North Dakota State University}%
\author{Theresa Elder}
\affiliation{Materials and Nanotechnology Program, North Dakota State University}%

\date{\today}

\begin{abstract}
Crumpling of a thin film leads to a unique stiff yet lightweight structure.  The stiffness has been attributed to a complex interplay between four basic elements - smooth bends, sharp folds, localized points (developable cones), and stretching ridges - yet rigorous models of the structure are not yet available.  In this letter we show that adhesion, the attraction between surfaces within the crumpled structure, is an important yet overlooked contributer to the overall strength of a crumpled film.  Specifically, we conduct experiments with two different polymer films and compare the role of plastic deformation, elastic deformation and adhesion in crumpling.  We use an empirical model to capture the behaviour quantitatively, and use the model to show that adhesion leads to an order of magnitude increase in ``effective'' modulus.  Going beyond statics, we additionally conduct force recovery experiments.  We show that once adhesion is accounted for, plastic and elastic crumpled films recover logarithmically.  The time constants measured through crumpling, interpreted with our model, show an identical distribution as do the base materials measured in more conventional geometries.
\end{abstract}

\pacs{Valid PACS appear here}
\maketitle

Thin films, a once passive part of design are quickly becoming the emphasis due to the emergence of thin film electronics and the demand for dynamic structures which are enabled through origami based design.\cite{Cao2008, Kim2012, Silverberg2014, Reis2015, Li2017}  These, and many other examples have lead to a resurgence in interest in the basic mechanics of thin structures, and more specifically in how energy becomes localized at singular points (developable cones) or in extended lines (folds).\cite{Cerda1999,Thiria2011}  How these localized objects interact with one another and with the materials in which they are created is still poorly understood, highlighted by the inability of models to accurately predict how a film resists deformation upon confinement.  In other words, it is still unclear why a crumpled ball is stiff, or if crumpled films differ at all from other low density structures like solid foams or lattice structures.  

To explain the stiffness in crumpling, several scaling models have been proposed which are based on the dominance of different physical features (ranging from the stretching in ridges that join two developable cones, the cost of completely collapsed folds or simply through dimensional analysis).\cite{Matan2002, Vlieg2006, Deboeuf2013}  Though often simple conceptually, the quantitative disagreement with experiment suggests that important physical features are being overlooked in current models.  For example, little thought has been given to the role of adhesion in crumpling despite the large amount of inter-sheet contact observed in confined thin films.\cite{Vlieg2006, Cambou2011}  The surface energy stored when two pieces of film come into contact can be estimated as $E_{s}\sim w R^{2}$, where $w$ is the work of adhesion (typically $10^{-2}$ N/m in polymer films) and $R$ is the size of the contact region.  The energy stored in bending a film into one, completely collapsed, fold scales as $E_{f}\sim E h^2 R$ where $E$ is the sheet's Young's modulus and $h$ its thickness.  The ratio of these two energies ($\delta=wR/Eh$) suggests that centimeter scale, micron thick films of modulus $10^{9}$~Pa are already dominated by adhesion.  With softer or thinner films the crossover scale would be much smaller.  In this letter we give direct experimental evidence that adhesion does play a significant role in the overall strength of a crumpled structure.  The presence of adhesion, however, does not affect the scaling relation between compaction force and confining dimension during crushing, implying that force-displacement measurements alone are inadequate for detecting the basic causes of stiffness in crumpled materials.

\begin{figure}
\centering
\includegraphics[width=\columnwidth]{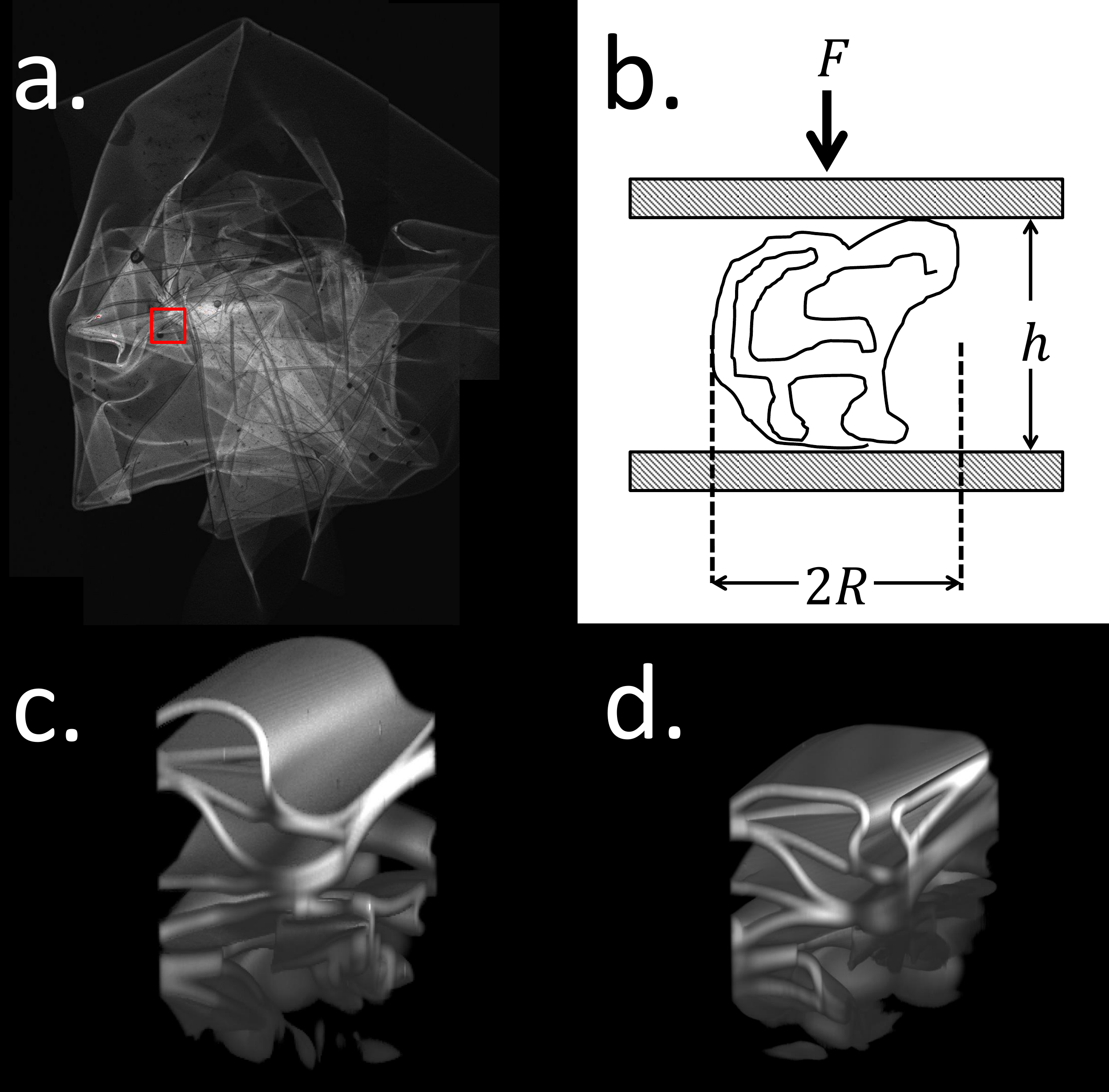}
\caption{a. A 2D image of a crumpled PDMS film in index matching fluid.  b.) a schematic showing the basic experimental geometry.  c.) 3D image at an early stage of compression. d.) The same location at a later stage of compression.}\label{Expt}
\end{figure}

Our experiments are conducted between two parallel glass plates, one of which is connected to a force transducer (Transducer Techniques, USA) and the other to a nanopositioner (Nexact, Physik Instrumente, Germany).  The setup is located under a confocal microscope in order to enable full 3D imaging of a film.  A sample is crumpled with tweezers and crushed in a quasi-static, displacement controlled experiment.  The mechanical setup and microscope are fixed to a standard air-floated optical table in order to minimize environmental noise.  Polycarbonate (PC) thin films were carefully created in house through spin-coating, flow-coating or casting of various concentrations of PC/chloroform solutions on freshly cleaved mica surfaces.  The resulting thicknesses range from $\sim 50$ nanometers to several millimeters.  Films were annealed at $\sim 180 ^\circ$C for approximately 1 hour to remove any internal stresses due the the fabrication processes.  Polydimethylsiloxane (PDMS) rubber films were created in a similar manner from Sylgard 184 mixed in the typical 10:1 prepolymer to crosslinker ratio.  Samples were crosslinked at a temperature of $\sim 80 ^\circ$C.  Nile red was incorporated into both materials in order to facilitate fluorescent imaging.  The basic setup, and typical microscope images can be seen in Fig.~\ref{Expt}, and full details can be found in~\cite{SM}.  

\begin{figure}
\centering
\includegraphics[width=\columnwidth]{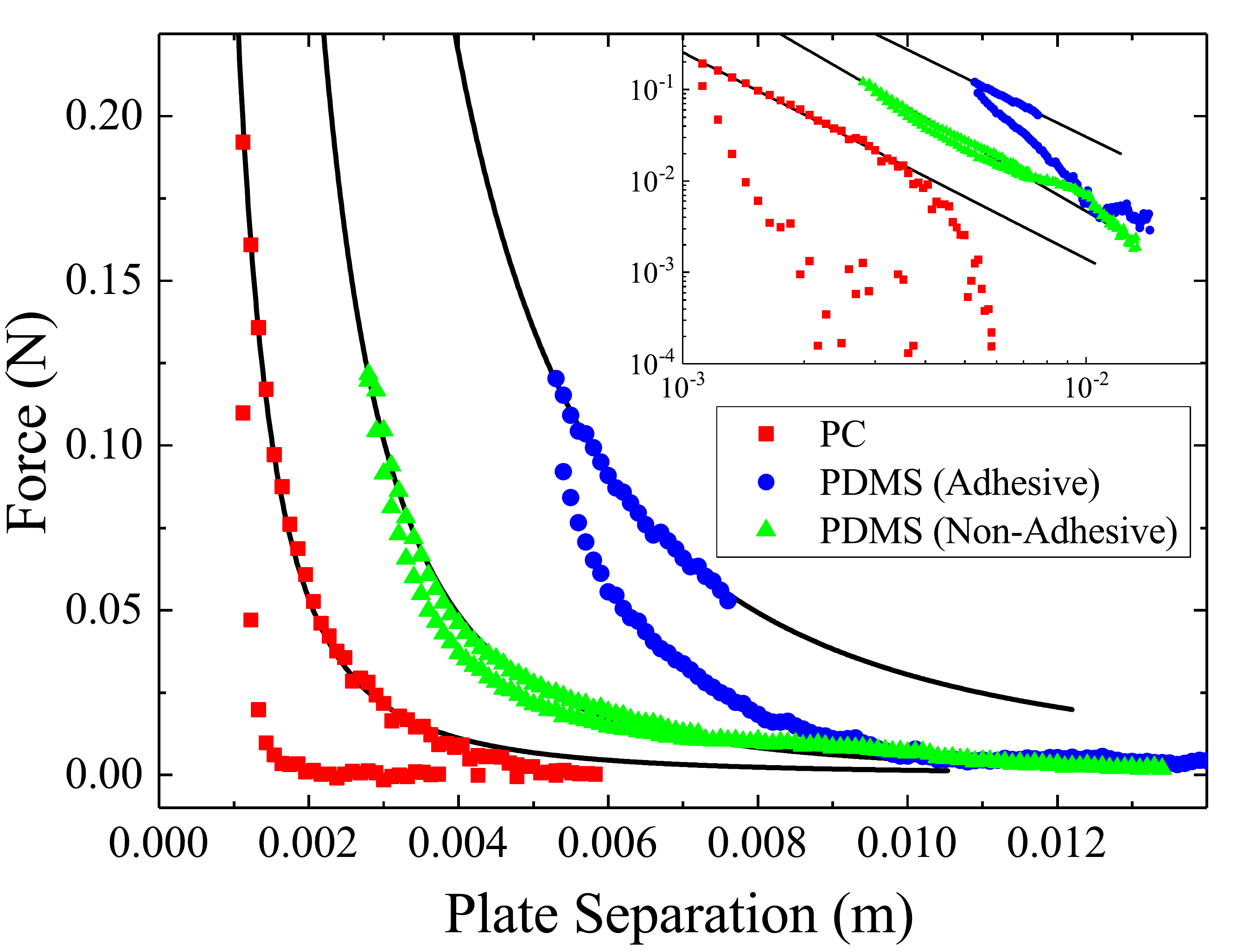}
\caption{Typical crumple compression data.  Red squares show data from a PC film with significant hysteresis ($16$~mm $\times$ $22$~mm, thickness $2$~$\mu$m).  Blue circles show a PDMS film which also has considerable hysteresis ($43$~mm $\times$ $40$~mm, thickness $86.5$~$\mu$m).  The same film is coated to reduce adhesion and retested (green triangles), showing a significant reduction in hysteresis.  Solid curves are power law fits to the compression step of the cycle.  Inset shows a log-log axis}\label{figdata}
\end{figure}

Key to this study is our ability to directly control the adhesion of the elastic PDMS film.\cite{Chaudhury1991}  The control is accomplished through the addition of a sparse, randomly oriented surface layer of either polystyrene colloids, or (more cost effectively) cornstarch.  The particles physically adhere to the PDMS surface and no additional adhesive agent is added.  The hard particle coating serves to keep the adhesive PDMS surfaces from coming into contact with themselves or the compression plates, while adding no possibility of hydrodynamic losses which can be significant for many lubricants confined to small gaps.  Conveniently, a particular PDMS film can be tested in both the adhesive \textit{and} the non-adhesive state, making the influence of adhesion quite clear.

Figure~\ref{figdata} shows typical force-displacement data from a crumpled PC film as well as the same PDMS film crumpled both in the adhesive (neat surface) and non-adhesive (coated surface) states.  The PC film shows clear hysteresis as is commonly observed in crumpling of plastic materials.  Here the energy loss is usually attributed purely to dissipation in plastic processes.  The PC film follows an apparent power law upon compression, and a different much steeper power law during retraction.  The adhesive PDMS film can also be quantified by a power law upon compression and shows a similarly large amount of hysteresis.  However, the hysteresis cannot be due to plasticity as the film is an elastomer; the loss is due to adhesion within the structure.  This point can be qualitatively proven by uncrumpling the film, coating its surface with a particle monolayer and then recrumpling and retesting.  In this non-adhesive case, the crumpling process is almost lossless, and once again shows a power law behaviour. 

The compressibility of a crumpled ball has been investigated by several researchers, and generally it is agreed that force-compaction ($F$ and $H$ respectively) experiments follow a power law scaling,
\begin{equation}\label{Powlaw}
F \sim F_{0} x^\alpha ,
\end{equation}
although there is little consensus on the origin of the exponent, $\alpha$, or the amplitude $F_{0}$.\cite{Matan2002, Deboeuf2013, Habibi2017, Balankin2011, Lin2008, Baimova2015}   Exponents ranging from $\sim 1.8$ to $15$ having been reported in materials ranging from polymer films to metal films and graphine.\cite{Matan2002, Deboeuf2013, Habibi2017, Balankin2011, Lin2008, Baimova2015}  Indeed, in the present work we report exponents of $2.8 \pm 0.5$ upon compression and $3.5 \pm 0.9$ upon retraction for PDMS and $7.7 \pm 5$ and $14.0 \pm 13.6$ for compression and retraction in PC.  The variability of the exponent cannot be ignored.

Matan et al. suggested the strength of the crumple had its origin in the forced stretching which occurs along the many ridges of film which join two adjacent developable cones (localized singular points of stretching which occur when a bent region is forced to bend in an orthogonal direction).\cite{Matan2002, Lobkovsky1995}  Using earlier scaling arguments for the energy stored elastically in a ridge, and neglecting any other interactions including self-avoidance of the sheet, the authors predicted that $F_{0} \sim Eh^{8/3}L^{16/3}R^{-10/3}$ and $\alpha \sim -8/3$ where $L$ is the film's lateral dimension, and $R$ the initial radius of the crumpled film.  Simulations were created in order to test this hypothesis with mixed results.  Vliegenthart and Gompper using a mesh of spring linked nodes and a dimensional argument, found an exponent of $\sim 14/9$ with phantom sheets (matching the ridge model), but a value of $\sim 2$ with more realistic self-avoiding sheets (which more closely matched the Matan et al. experiments).\cite{Vlieg2006}  It is interesting that the exponents measured for the simulated self-avoiding sheet, when input into the dimensional scaling, implies only a single bend is present (scaling as $F\sim Eh^3 L /H^2$).\cite{Elder2018}

The exponents reported in this letter do not agree with the existing predictions.  The exponents measured for the PDMS films (adhesive and non-adhesive) do closely match the ridge-model exponent of $-8/3$, however, comparing the amplitude ($F_0$) yields only weak correlation and several orders of magnitude error in scale (see \cite{SM}).  The disagreement in the PDMS data may be due to the lower F{\"o}ppl-von K{\'a}rm{\'a}n numbers accessed by the experiments ($\gamma\sim L^2/h^2\sim10^4 -10^7$), as the asymptotic scaling on which it is based is only valid above $\gamma =10^8$.  The PC films are well within the asymptotic limit ($\gamma_{PC}\sim 10^5-10^{12}$), but give exponents that are far too large.  $F_{0}$ is also off by several orders of magnitude for both PDMS and PC.

The clear inconsistencies of the early crumpling models motivated more recent models which have focused on differing physics.  A model based on energy storage in the irrecoverable plasticity occurring as curvature is localized into sharp folds was developed by Deboeuf et al. in \cite{Deboeuf2013}.  The model predicts several different values for $\alpha$, ranging from 1 to 4, depending on the underlying structure and type of compression.  Additionally the model predicts an amplitude of $F_{0}=Eh^{2} L^\alpha$.  The model has been validated experimentally through the crushing of cylindrically bent (rolled-up) sheets and sheets confined in 3D by a wire mesh.  Exponents were found to depend on geometry and material properties.\cite{Deboeuf2013, Habibi2017}  The fold model applied to our data shows qualitative agreement, but once again, a quantitative error of several orders of magnitude (see~\cite{SM}).  Furthermore, the exponents measured for PC are beyond what is expected in this model.  

We find a small modification of this scaling,
\begin{equation}\label{empfit}
F=Eh^2 \bigg{(} \frac{2R_{0}}{H} \bigg{)} ^{\alpha}
\end{equation}
where $L$ is replaced with $2R_{0}$ the initial size of the crumple, is \textit{quantitative} in its fit to our data.    Here we probe the existing network, not how the network was constructed which justifies the use of $R$ over $L$.  The exponent is then related to the network structure of ridgid elements, not all of which are initially load-bearing.\cite{networks}  The small, empirically motivated, modification we suggest would not significantly alter conclusions of earlier experiments (in a cylindrical film $R=L$) but, importantly, the change allows us to more deeply understand the role of adhesion in crumpling.

\begin{figure}
\centering
\includegraphics[width=\columnwidth]{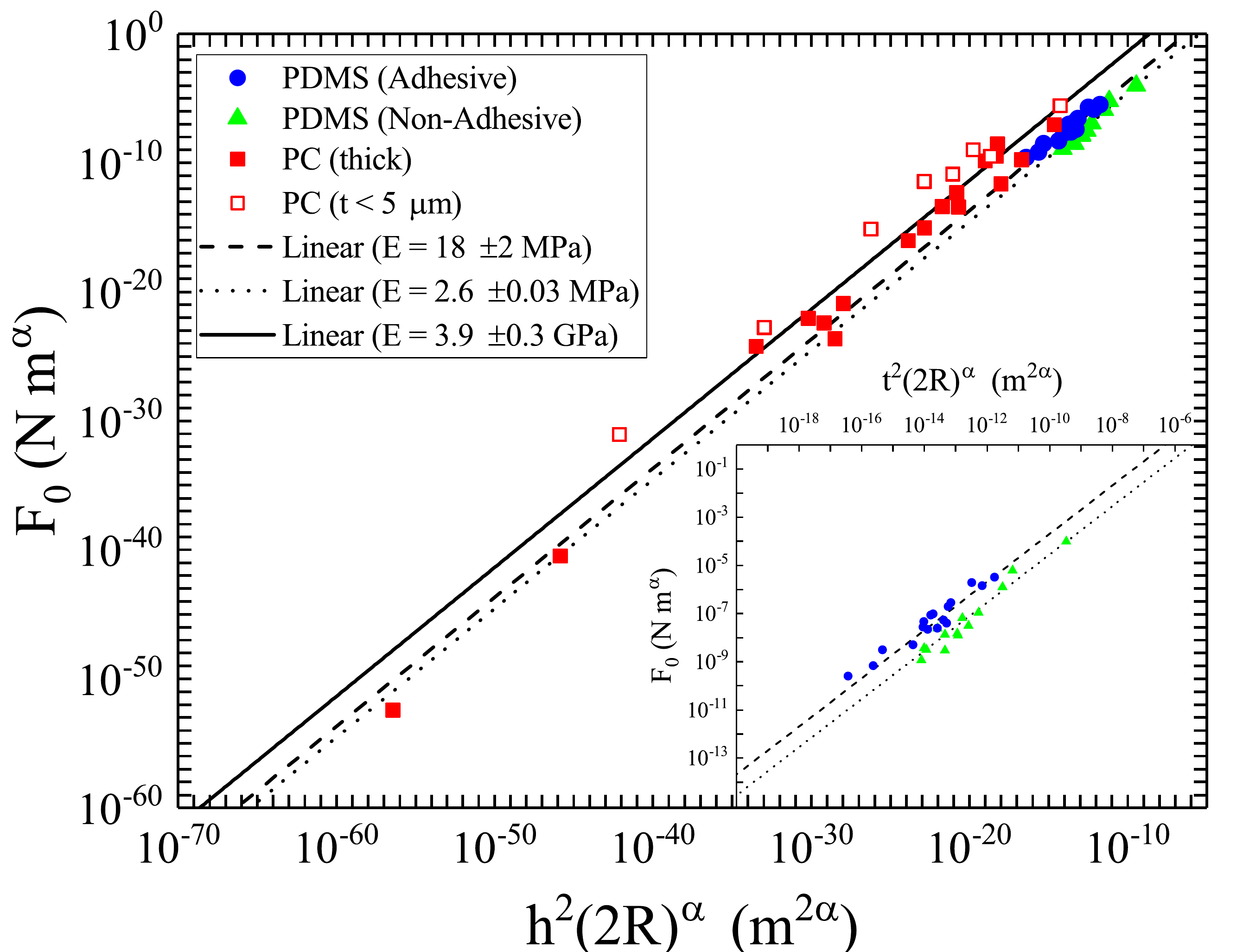}
\caption{Force amplitude plotted against $h^2R^\alpha$.  All data is well fit by Eqn.~\ref{empfit}, yielding a measured value for Young's modulus for each system.  The measured modulus of Non-adhesive PDMS and PC match their bulk values quite closely, however, the adhesive PDMS shows an `effective' Young's modulus an order of magnitude larger than is expected.}\label{figamp}
\end{figure}

Figure~\ref{figamp} shows a plot of $F_{0}$ vs $h^{2} (2R)^{\alpha}$ for each set of crumple data.  The plot shows a clear linear relationship (over 50 orders of magnitude), the slope of which can be identified as the film's modulus.  Quantitatively, slopes of $3.9\pm0.3$~GPa (PC), and $2.6\pm 0.03$~MPa (non-adhesive PDMS) are found, which can be compared to the independently measured values of $1.6$~GPa and $1.7$~MPa for PC and PDMS respectively.\cite{Elder2018}  The adhesive PDMS shows an order of magnitude discrepancy; it's `effective' Young's modulus at $18\pm2$~MPa is an order of magnitude too large.  Adhesive interactions are also present in the PC films, however, they are masked by the high modulus of the film.  The high modulus decreases the amount of true interfacial contact between two interacting segments of film, and the overall gain in contact energy is insignificant.  However, as the thickness is reduced, even PC films will be deformed by surface forces creating high amounts of true surface contact.  If the PC data is separated into `thick' (filled squares, $h>5$~$\mu$m) and `thin' (open squares, $h<5$~$\mu$m) samples, adhesion is once again apparent.  Even rigid films are affected by adhesion, particularly at high $\gamma$ values.

The validity of Eqn.~\ref{empfit} can be further explored through analysis of the dynamics of the crumpled structures.  Existing studies have shown varied results; some showing logarithmic \cite{Matan2002, Deboeuf2013, Habibi2017, Balankin2011, Lin2008, Baimova2015} while others show stretched exponential \cite{Albu2002} dynamics.  Simulations have not yet been able to track dynamics on realistic timescales and there has been little attempt to explain the behaviour theoretically.  If Eqn.~\ref{empfit} is valid, it suggests a possible solution as it allows firm prediction for the origin of any observed dynamics.  In a force recovery experiment all variables in Eqn.~\ref{empfit} are fixed, leaving only the modulus as a function of time.  This hypothesis is easily tested by comparing the modulus measured in a crumple, with the modulus measured in a more direct geometry.

\begin{figure}
\centering
\includegraphics[width=\columnwidth]{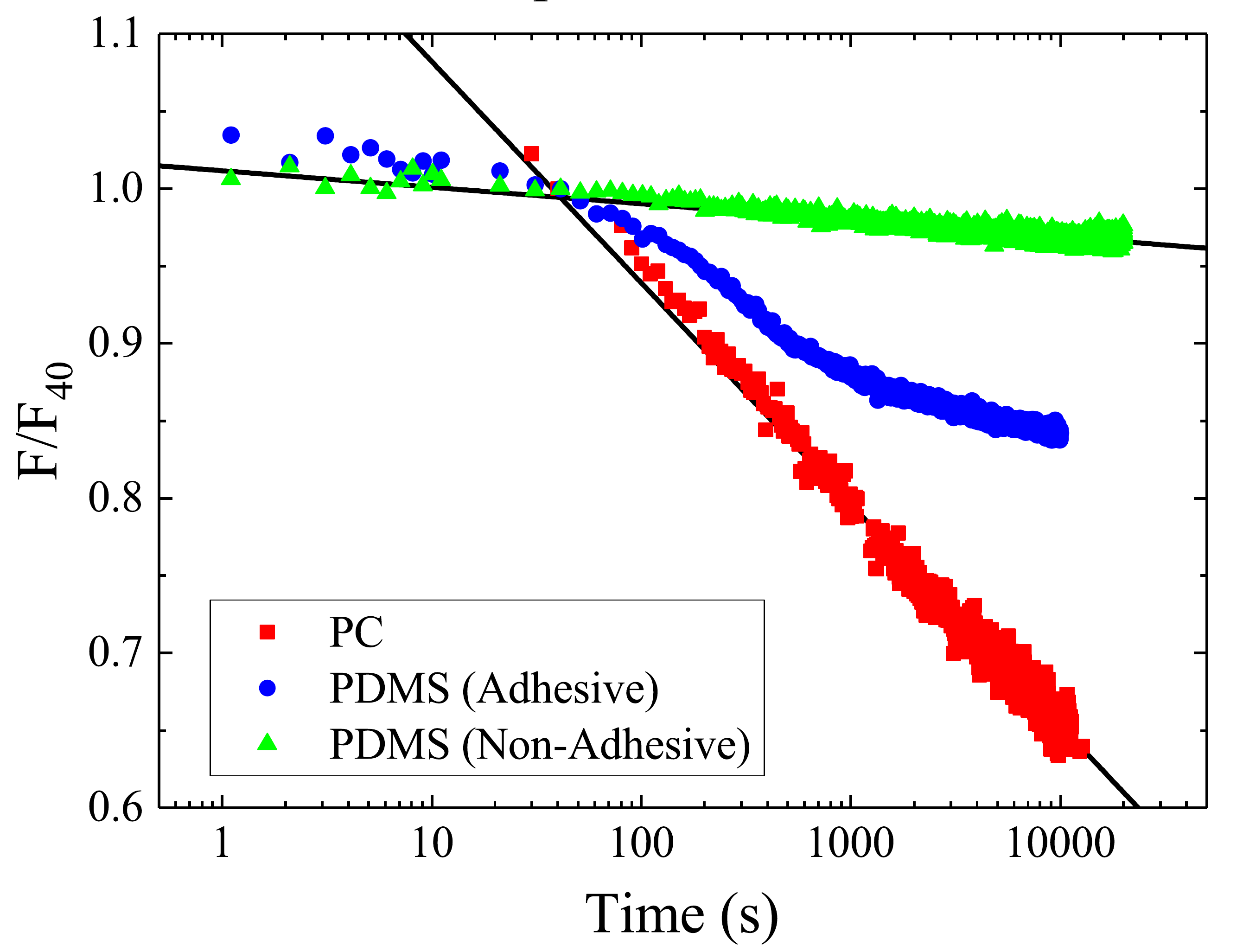}
\caption{Force recovery data from the same samples shown in Fig.~\ref{figdata}.  All curves are normalized by the instantaneous force recorded at 40~s.  Logarithmic fits are shown in black.}\label{FRdata}
\end{figure}

Force recovery experiments were conducted by stopping each indentation cycle at a predetermined plate separation and subsequently monitoring force as a function of time.  Figure~\ref{FRdata} shows typical data from the PC, adhesive PDMS and non-adhesive PDMS systems normalized by their instantaneous force at $40$~s.  A time of $40$ seconds was chosen because each data set shows a shoulder at shorter times, which reflects the history of the sample as it slowly approaches the point of force recovery.  At times greater than $40$~s the PC data shows a clear logarithmic decay which is well fit with the function: 
\begin{equation}\label{loggo}
F/F_{40} = \beta \log {t} + B
\end{equation}
where $t$ is time, $B$ is a constant approximately equal to 1 and $\beta$ is the relevant relaxation constant.  Relaxation constants for PC were found to vary from sample to sample which we discuss further below.

PDMS in its natural (adhesive) state shows a similar decay in force over time, however, it reaches an inflection point and begins to asymptotically approach a constant value.  Notably, the shape is consistent with a stretched exponential trend.  The arrest in the dynamics, however, can be directly ascribed to the adhesion present in the system by comparison with PDMS films in the non-adhesive state.  In the non-adhesive state, data shows a smooth logarithmic trend, similar to the PC films, and is once again fit with Eqn.~\ref{FRdata}.  Variation in $\beta$ was also noted.

\begin{figure}
\centering
\includegraphics[width=\columnwidth]{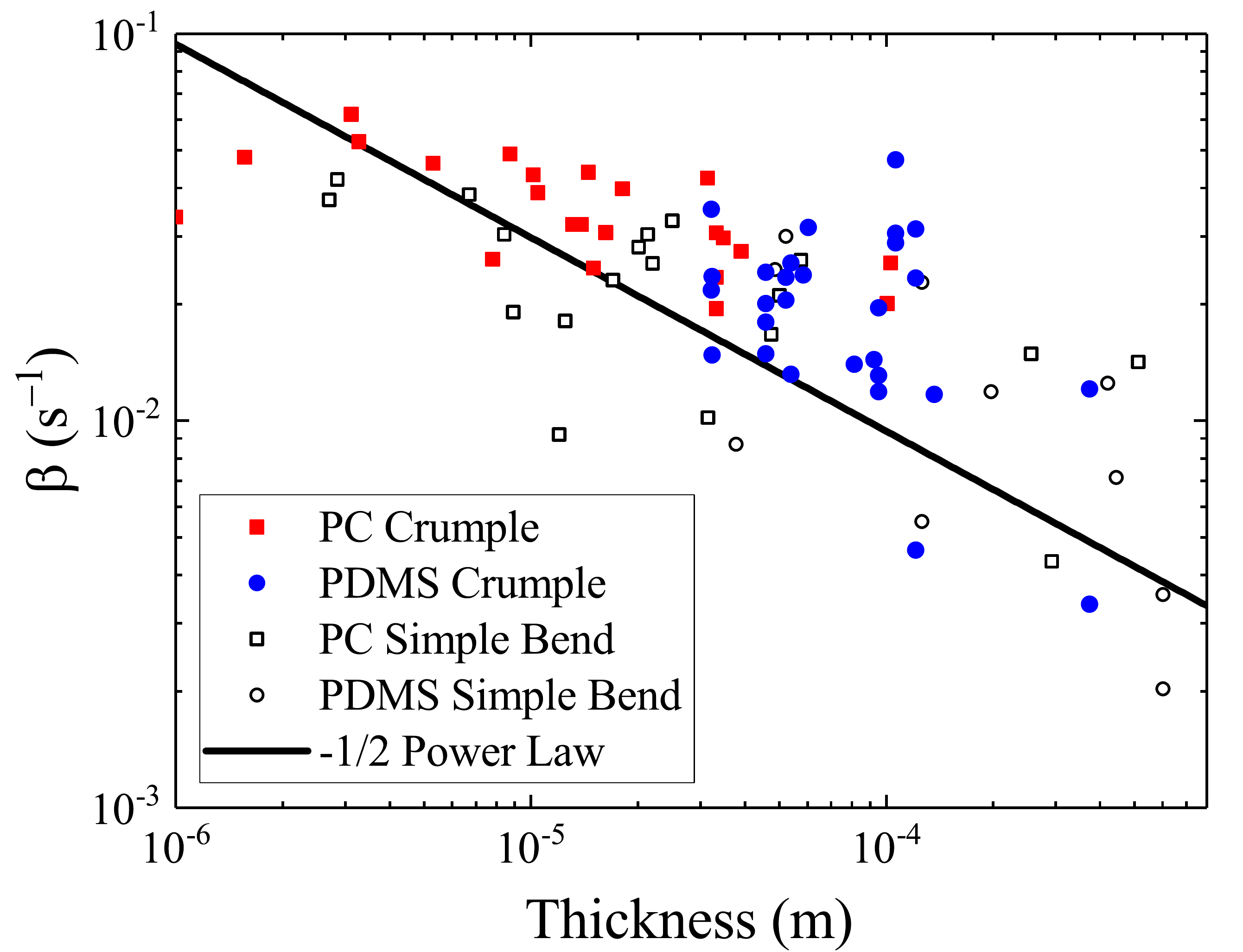}
\caption{The dependence of the logarithmic time constant on film thickness.  Solid symbols show values for PC (squares) or non-adhesive PDMS (circles).  Open symbols represent the results of force recovery measurements of simply bent PC(squares) or PDMS (circles), data reproduced from \cite{Elder2018}.}\label{relaxation}
\end{figure}

Fig.~\ref{relaxation} shows the relaxation constants measured in several experiments with PC or non-adhesive PDMS as a function of film thickness (solid symbols).  The data shows  some variation, but is reasonably fit by a square root relationship which is an indicator of diffusive relaxation.  More importantly, the data measured in the crumpled state perfectly overlaps force recovery data taken in a much simpler geometry.  In \cite{Elder} thin PC and PDMS films were curved into a single bend and force recovery experiments were conducted.  The single bend experiments showed logarithmic trends, and similar relaxation constants were measured which we reproduce here in Fig.~\ref{relaxation} (hollow symbols).\cite{Elder2018}  The near perfect overlap between the current crumpling experiments and the earlier single bend experiments indicates that all geometric details have been accounted for.  Eqn.~\ref{empfit} has allowed the recovery of a true material property (Young's modulus) from the complex crumpled geometry.  

In conclusion, we have directly explored the role of adhesion in crumpled polymer films.  We use a simple, empirical model to fit the observed power-law force data \textit{quantitatively} and use the fit to reveal an order of magnitude increase in the effective modulus of self-adhesive films.  Our interpretation is strengthened by force recovery experiments which show differences between adhesive and non-adhesive films.  In the absence of adhesion we show definitively that force recovery in crumpled structures is due only to material properties.

The authors gratefully acknowledge that this work was supported by the Air Force Office of Scientific Research (AFOSR) under Grant FA9550-15-1-0168.

\bibliography{crumple}

\end{document}


\preprint{APS/123-QED}

\title{Supplement to The Roll of Adhesion in the Mechanics of Crumpled Polymer Films}

\author{Andrew B. Croll}
\email{andrew.croll@ndsu.edu}
\affiliation{Department of Physics and Materials and Nanotechnology Program, North Dakota State University}%
\author{Timothy Twohig}
\affiliation{Department of Physics, North Dakota State University}%
\author{Theresa Elder}
\affiliation{Materials and Nanotechnology Program, North Dakota State University}%

\date{\today}


\maketitle

\section{Additional Sample Preparation Details}
\subsection{PC}
Polycarbonate (PC) was used as received from Scientific Polymer Products and was reported to have a molecular weight of 60 000 Daltons.  Solutions were created by dissolving the polymer in chloroform (Fisher Scientific, Optima grade) to various weight percents up to 10\%.  Nile red, a fluorescent dye, was often added to solution in trace amounts.  Films were created in several ways.  Below $\sim 2$ microns in thickness, films were created by spin coating solution on freshly cleaved mica supports.  Instabilities limit the thickness in this case.  Larger thicknesses were created through drop casting polymer solution on freshly cleaved mica supports in a chloroform saturated environment, which was allowed to slowly evapourate over several days.  Drop casting was limited to thicknesses above $\sim 2$~$\mu$m due to dewetting instabilities which occurred during the casting process.

After creation, films were annealed for $\sim 1$~h at a temperature of 453~K in order to remove any residual stress caused by the sample preparation techniques.  Films were scored with a scalpel blade, then released on a Milli-Q water surface.  Film thickness was measured with atomic force microscopy (the thinnest samples) or was measured with confocal microscopy (the thicker samples).  Each film was measured in several locations and an average thickness was used.  Variation was typically 12\%.

\subsection{PDMS}
Elastomeric polydymethylsiloxane (PDMS) films were created with sylgard 184 (Dow Corning) mixed in a 10:1 ratio.  Films were cast on glass slides which were covered in a thin layer of poly(acrylic acid) through spin coating a 5\% by weight water/poly(acrylic acid) solution.  PDMS mixtures were degassed in a vaccuum oven, then coated through drop casting, or spin coating on a poly(acrylic acid) coated glass slide.  Films were then placed in the vaccuum oven and annealed for 1~h at 353~K under vacuum.  Films were cooled to room temperature before use.

Films were scored with a scalpel blade, and released on a Milli-Q water surface.  Films were removed from the water surface, dried, and immersed in a toluene Nile Red solution.  After a short time ($\sim 10$~min) films were removed from the toluene solution, excess solution was removed from the films surface and the film was allowed to dry.  Once dry, films were stored in the flat state for 24 hours before use.  Thicknesses were measured in the same manner as the PC films outlined above.

\section{Additional Measurement Details}
\subsection{Crumple Radius}
The setup used in this experiment has no bounding walls in the horizontal direction.  This means that the definition of the lateral extent of the crumple ($R$) is ambiguous without further clarification.  We adopted a definition based on the average lateral extent of the crumple as measured in images taken on two orthogonal axis through the side of the crumple, or via a 2D projection of the crumple taken from the top with the confocal microscope.  The side images were used preferentially as  a smallest radius and largest radius in the images could easily be defined.  Both smallest and largest radii were measured independently twice, along two orthogonal axis and averaged to a single value for $R$.  Each image was calibrated independently ensuring no drift in scale.

The largest radius did not change significantly during compaction for any of the materials studied.  The smallest radius did often grow during compaction, indicating a densification of the structural network occurs during crushing of at least some of the crumples studied.  No correlation was noted between the rate of growth of the smallest radius and scaling exponent($\alpha$ in the text), force amplitude ($F_{0}$ in the text) or time time constant.

\begin{figure}
\centering
\includegraphics[width=\columnwidth]{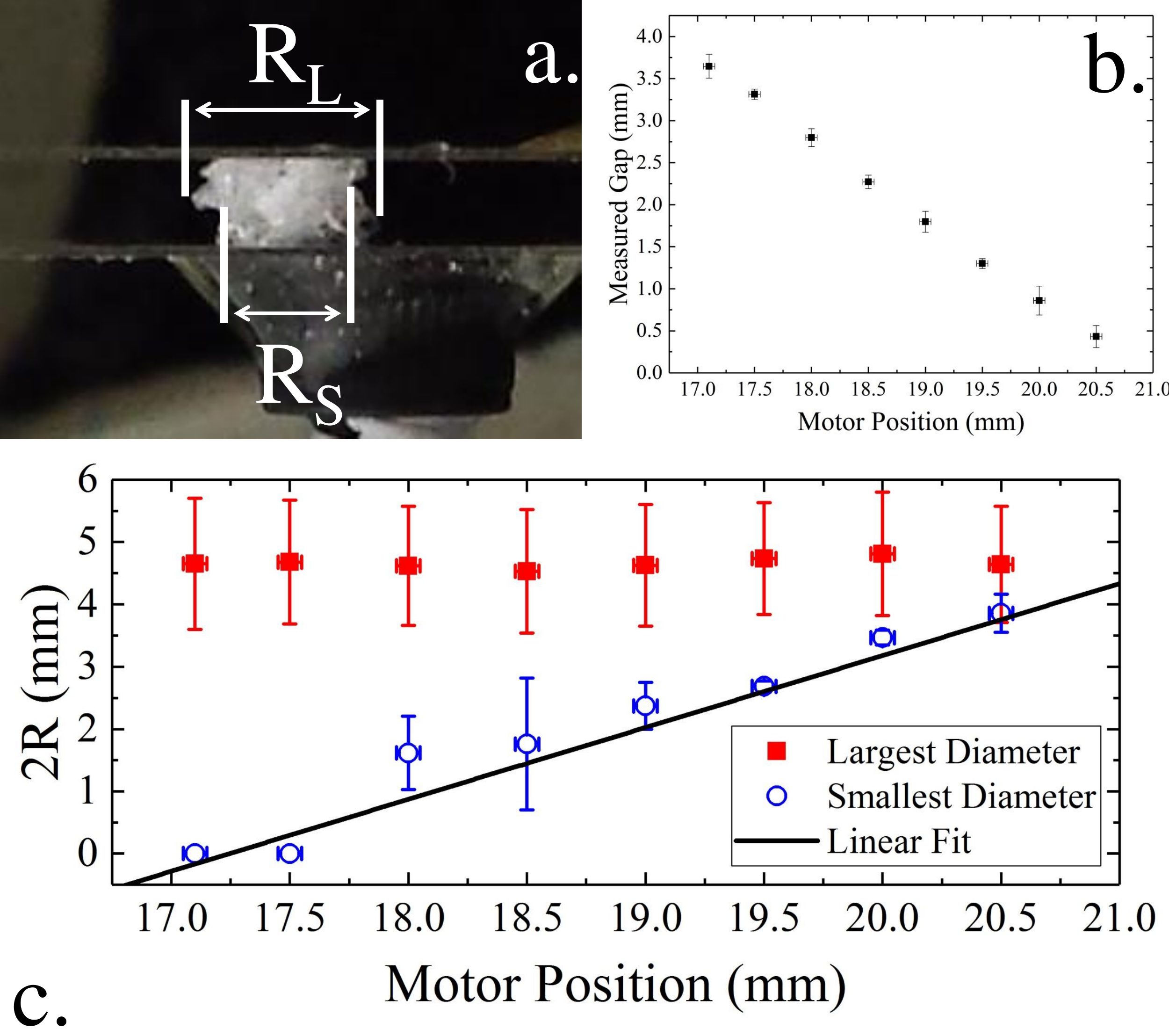}
\caption{a. A typical image of the horizontal extent of a crumple.  In this case, a PC film of diameter $4.7$~mm is shown.  $R_{L}$ and $R_{S}$ refer to the largest radius and smallest radius observed in the image.  b.) Plate separation as a function of motor position showing a linear trend indicating the apparatus is still stiff compared to the crumpled film.  c.) Change in the largest (solid squares) and smallest diameter (open circles) as a function of compression.  The larger diameter does not change during the experiment, however the smaller diameter does grow in an approximately linear manner (a slope of 1.15 was found with a linear fit shown as the solid line)}\label{Expt}
\end{figure}

\subsection{F{\"o}ppl-von K{\'a}rm{\'a}n Number}
It is not uncommon to classify mechanical experiments with thin films by their range of F{\"o}ppl-von K{\'a}rm{\'a}n numbers (FvK), defined as $\gamma = L^2/h^2$, where $L$ is the lateral size of a film and $h$ its thickness.  For example, scaling behaviour is often limited to a certain range of FvK numbers.  In the present work, we characterized the force required to compress a crumpled film by an amplitude ($F_{0}$) and a scaling exponent ($\alpha$) related through the power law: $F=F_{0} x ^\alpha$, with $F$ the applied force and $x$ the gap size.  Given the variation we observed in both quantities, it was natural to check for correlation with the FvK number to see if the variation of our experiments was simply due to our approach to an asymptotic limit.

\begin{figure}
\centering
\includegraphics[width=\columnwidth]{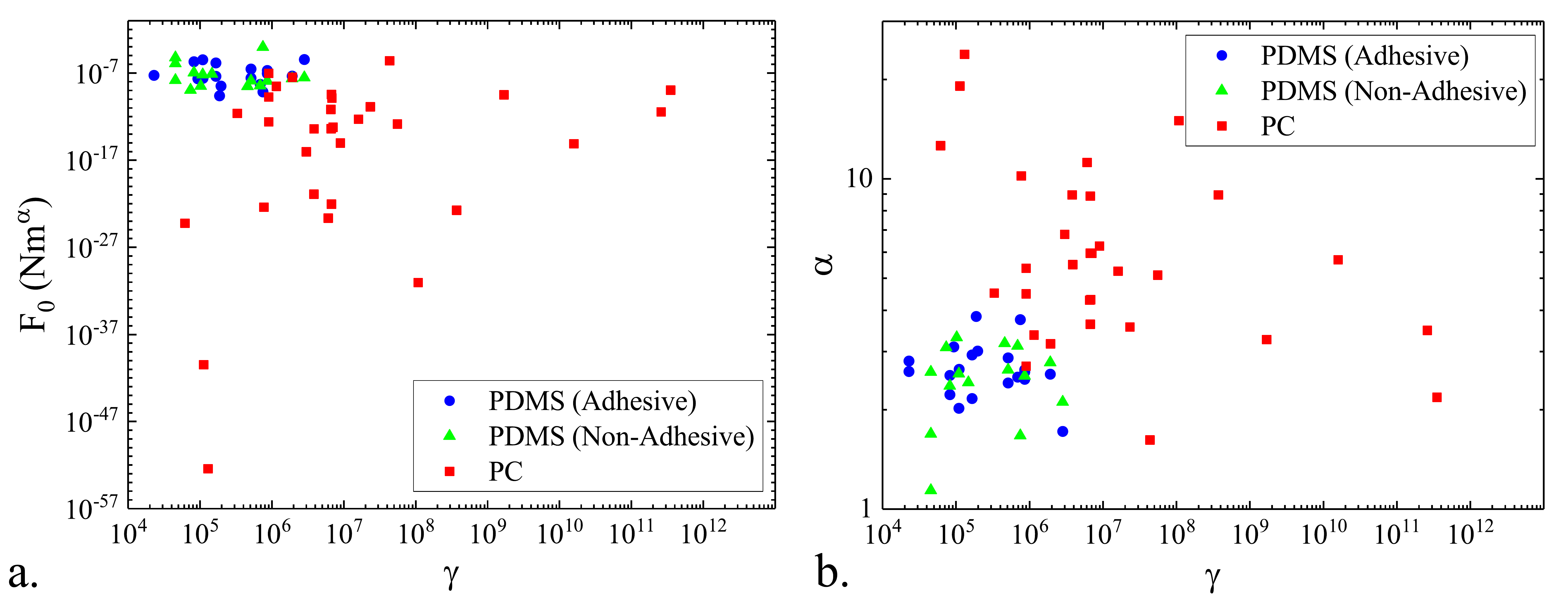}
\caption{Relation between the F{\"o}ppl-von K{\'a}rm{\'a}n number, $\gamma$, and the amplitude (a.) and the scaling exponent (b.) of power law fits to compressed crumples in several materials.  Data appears uncorrelated in the range available to experiments.}\label{FVK}
\end{figure}

Figure~\ref{FVK} shows a clear lack of correlation between power law amplitude (a.) or exponent (b.) and the F{\"o}ppl-von K{\'a}rm{\'a}n number.  For example, as the FvK number increases we see no signs of the measurements approaching a constant value.  We note our experiments are limited to only about 8 orders of magnitude in FvK number, which could (in principle) hide any trends occurring at higher FvK numbers.

\section{detailed comparison with models}
\subsection{Ridge Model}

Matan and coworkers proposed a scaling model to describe the compaction of a crumpled film between two walls in \cite{Matan2002}.  The model considered the stretching incurred in a ridge joining two developable cones as the primary energetic cost in the system, and estimated the number and size of ridges from the density of the crumpled film.  Specifically, the model predicts compaction force to scale as:
\begin{equation}\label{witten}
F \sim Eh^{8/3}L^{16/3}R^{-10/3}x^{-8/3},
\end{equation}
where $E$ is Young's Modulus, $h$ is film thickness, $L$ is the lateral extent of the film, $R$ is the crumple's radius and $x$ is the gap between the compressing walls.  As noted in the manuscript, we find exponents similar to $8/3$ only in the PDMS films suggesting the argument is plausible for elastic materials.  However, it is important check the scaling of the force with film dimensions and crumple radius as these are also prominent features of the model.  If all force-compression data is fit with a power law ($F_{0}$ if $F=F_{0} x^{\alpha}$) then the remaining variables can examined in a plot of $F_{0}$ as a function of $h$, $L$ and $R$.  

\begin{figure}
\centering
\includegraphics[width=\columnwidth]{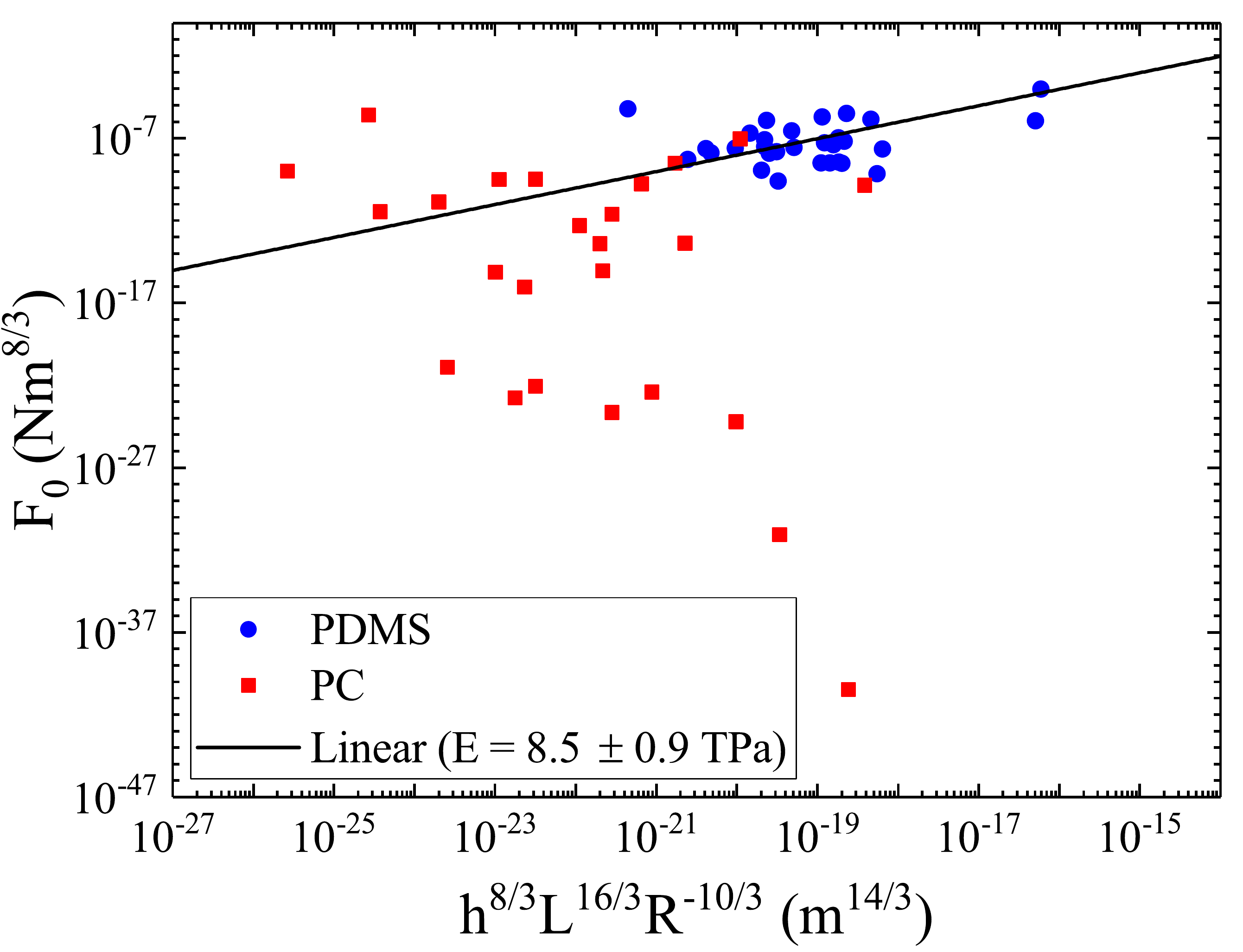}
\caption{Scaling prediction of the Matan et al. model.  The large scatter in the data makes it difficult to confirm the linear fit with either material, though it is plausible for the PDMS data.  Quantitatively, the PDMS data predicts a modulus of 8.5 TPa, which is many orders of magnitude from the known modulus of ~2 MPa.  Such disagreement is beyond any reasonable factors neglected in Eqn.~\ref{witten}}\label{wittenfig}
\end{figure}

Figure~\ref{wittenfig} shows the force amplitude of both PDMS and PC data.  The PC data shows considerable scatter, and makes it unlikely that a linear fit is reasonable.  This is consistent with the high, and variable, scaling exponent measured with respect to the confining dimension.  Both observations are clear signs that PC is not well described by Eqn.~\ref{witten}.  The PDMS data of Fig.~\ref{wittenfig} shows scatter, but is plausibly consistent with a linear trend (shown as a solid line).  On this axis, the slope of the liner fit is interpreted as the film's modulus and for the PDMS shown yields a modulus of $8.5 \pm 0.9$~TPa, far beyond what is reasonable for PDMS.  Once again it must be concluded that Eqn.~\ref{witten} is not, in fact, consistent with the data.

\subsection{Fold Model}
\begin{figure}
\centering
\includegraphics[width=\columnwidth]{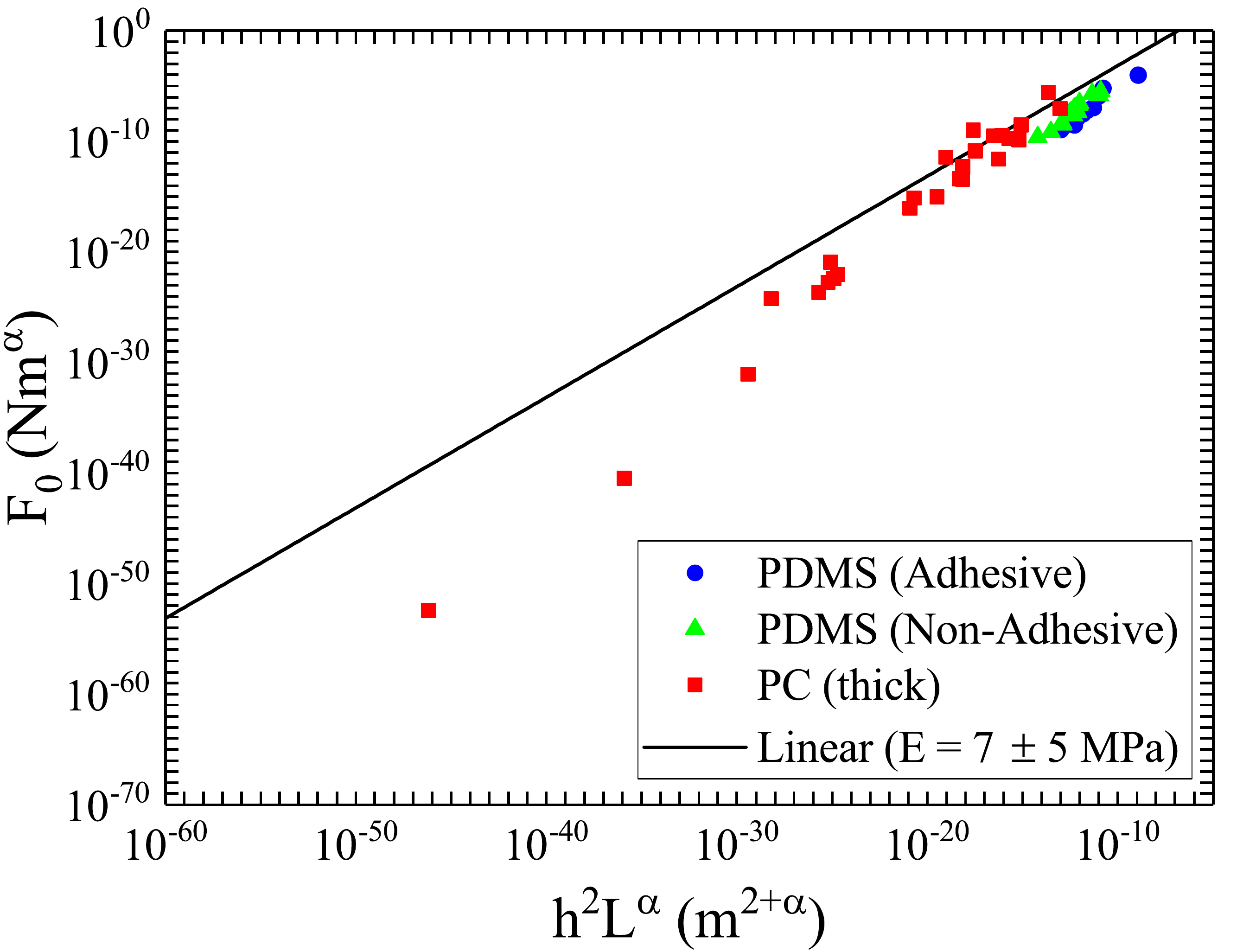}
\caption{Scaling predicted by fold model.  A linear fit to the PC data is shown as the solid line, resulting in a modulus of 7 MPa far below what is expected.}\label{fold}
\end{figure}

Deboeuf et al. proposed an alternative model based on the dominance of the energetic cost of single, fully collapsed ridges (e.g. curvature $\rightarrow 1/h$) in crumpling a film.\cite{Deboeuf2013}  The model predicts compressive force to scale as $F\sim E h^2 (x/L)^{- \alpha}$, where alpha depends on the exact type of folding but ranges from 1 to 4.  Figure~\ref{fold} shows $F_{0}$ as measured for PDMS and PC as a function of $h^2 L^\alpha$.  A linear fit to the PC data is also shown in the figure.  As in Fig.~\ref{wittenfig}, consistency with a linear correlation means the predicted scaling is accurate and the slope is the modulus up to a scaling constant.  However, we observe the correlation to be imperfect and the modulus to be inaccurate by several orders of magnitude (we find a modulus of 7 MPa for PC).  The PDMS data ranges only a few orders of magnitude but has a similar error in correlation.  More importantly, 3D imaging shows very few fully collapsed folds in either material - counter to the foundational assumption of the model.  PDMS, in fact, rarely showed any sizable regions of curvature $\rightarrow 1/h$.  Additionally, the reader is reminded of the high scaling exponent observed for PC which is also inconsistent with the predictions of the fold model.  We conclude that the scaling predicted by the fold model appears close to what is found in experiment, but the model itself is not fully consistent with observations.

\section{Acknowledgments}
The authors gratefully acknowledge that this work was supported by the Air Force Office of Scientific Research (AFOSR) under Grant FA9550-15-1-0168.

\bibliography{crumple}